\documentclass[aps,pra,twocolumn,amsmath,amssymb,bibliography,superscriptaddress]{revtex4-2}
\usepackage{times}
\usepackage[english]{babel}
\usepackage{latexsym}
\usepackage{graphics}
\usepackage{subfigure}
\usepackage{epsfig}
\usepackage{color}
\usepackage{hyperref}
\usepackage{braket} 
\usepackage{amstext}
\usepackage{amssymb}
\usepackage{graphicx}
\usepackage{esint}
\usepackage{braket}
\usepackage{babel}
\usepackage{amsmath}
\usepackage{comment}
\usepackage{dsfont}
\usepackage{xcolor}
\definecolor{resultblue}{RGB}{30, 90, 160}
\hypersetup{
colorlinks=true,
citecolor=resultblue,
linkcolor=resultblue,
urlcolor=resultblue
}

\makeatletter
\renewcommand{\fnum@figure}{FIG.~\thefigure}
\renewcommand{\fnum@table}{TABLE~\thetable}
\makeatother

\footnotetext{These authors contributed equally to this work.}

\begin{document}
\title{
Mitigating quantum decoherence via global optimal control
}

\author{A. Abedi$^{\S}$}
\email{aabedi@planckian.co}
\affiliation{Planckian srl, I-56127 Pisa, Italy}
\affiliation{NEST, Scuola Normale Superiore, I-56127 Pisa, Italy}
\author{R. Menta$^{\S}$}
\email{rmenta@planckian.co}
\affiliation{Planckian srl, I-56127 Pisa, Italy}
\affiliation{NEST, Scuola Normale Superiore, I-56127 Pisa, Italy}
\author{J. Despres$^{\S}$}
\author{M. Polini}
\affiliation{Planckian srl, I-56127 Pisa, Italy}
\affiliation{Dipartimento di Fisica dell’Universit\`{a} di Pisa, Largo Bruno Pontecorvo 3, I-56127 Pisa, Italy}
\author{F. Caravelli}
\affiliation{Planckian srl, I-56127 Pisa, Italy}
\author{V. Giovannetti}
\affiliation{Planckian srl, I-56127 Pisa, Italy}
\affiliation{NEST, Scuola Normale Superiore, I-56127 Pisa, Italy}

\begin{abstract}
We show that global optimal control can drastically suppress the impact of decoherence in globally driven superconducting quantum computing architectures, taking as a prototype a recently proposed quasi-two-dimensional ladder geometry. Using a tensor-network-based approach, we quantify how amplitude-damping and dephasing channels degrade the flow of quantum information along the ladder and the fidelity of one- and two-qubit gate operations. We then demonstrate that shaping the global drive compresses the gate sequences by an order of magnitude in time, restoring high gate fidelities. We stress that this mitigation is far from trivial: in a globally driven processor, dissipation acts on every physical qubit---including those outside the logical register that sustain the surrounding ordered phases---so its impact cannot be suppressed by protecting an isolated subsystem, and is instead overcome purely through the temporal shaping of the global drive.
\end{abstract}

\maketitle

\section{Introduction}
\label{sec:intro}

Over the last two decades, remarkable experimental progress in the control of
quantum lattice models has provided strong momentum to the fields of quantum
simulation and quantum computing~\cite{Lloyd_universality_1996}. However, the latter requires scalable
architectures capable of controlling large qubit arrays while maintaining high
fidelity for every quantum operation. Among the available experimental
platforms---including photonic systems, trapped ions, neutral Rydberg atoms,
and coherent states of light---superconducting qubits stand out for their
scalability and compatibility with existing nanofabrication
techniques~\cite{Martinis2019quantum,supremacy_PRL_2021,bravyi2022future,
ezratty2023perspective,vandamme2024}. Traditionally, superconducting
architectures rely on individually addressable control lines to interact
locally with each qubit. Although this approach generally reduces the total
number of quantum operations required, its drawback lies in the introduction
of substantial wiring complexity, thermal noise, and cross-talk
errors~\cite{krinner2019,Girvin2008,Gambetta2017}. As larger quantum
processors are considered, these disadvantages, becoming increasingly pronounced,
require innovative strategies for efficient and reliable qubit control~\cite{rosenberg2017,Kjaergaard2020,Mohseni2024, Marvin2026}.

A promising alternative consists of globally-controlled architectures in
which a small number of shared control lines act simultaneously on entire
subsets of qubits, thereby reducing the complexity of the control electronics
and simplifying large-scale implementations. Such schemes were initially
proposed by Lloyd~\cite{Lloyd_1993} and subsequently studied in
depth by Benjamin et al.~\cite{benjamin_2001_1,benjamin_2001_2,Levy_2002,benjamin_2003,benjamin-bose_2004,Ivanyos_2005,Kay_2004,Fitzsimons_2006,
Silva_2009}. Despite their considerable theoretical promise, these proposals
long remained without concrete quantum-technology implementations. More
recently, globally-controlled schemes have been actively investigated across
several physical platforms, including Rydberg
atoms~\cite{cesa2023universal}, spin
qubits~\cite{Patomaki_2024, Benjamin2026}, and superconducting
circuits~\cite{menta2024globally,cioni2024conveyorbelt,menta2025building}.

\begin{figure}[t]
\centering
\includegraphics[width=\columnwidth]{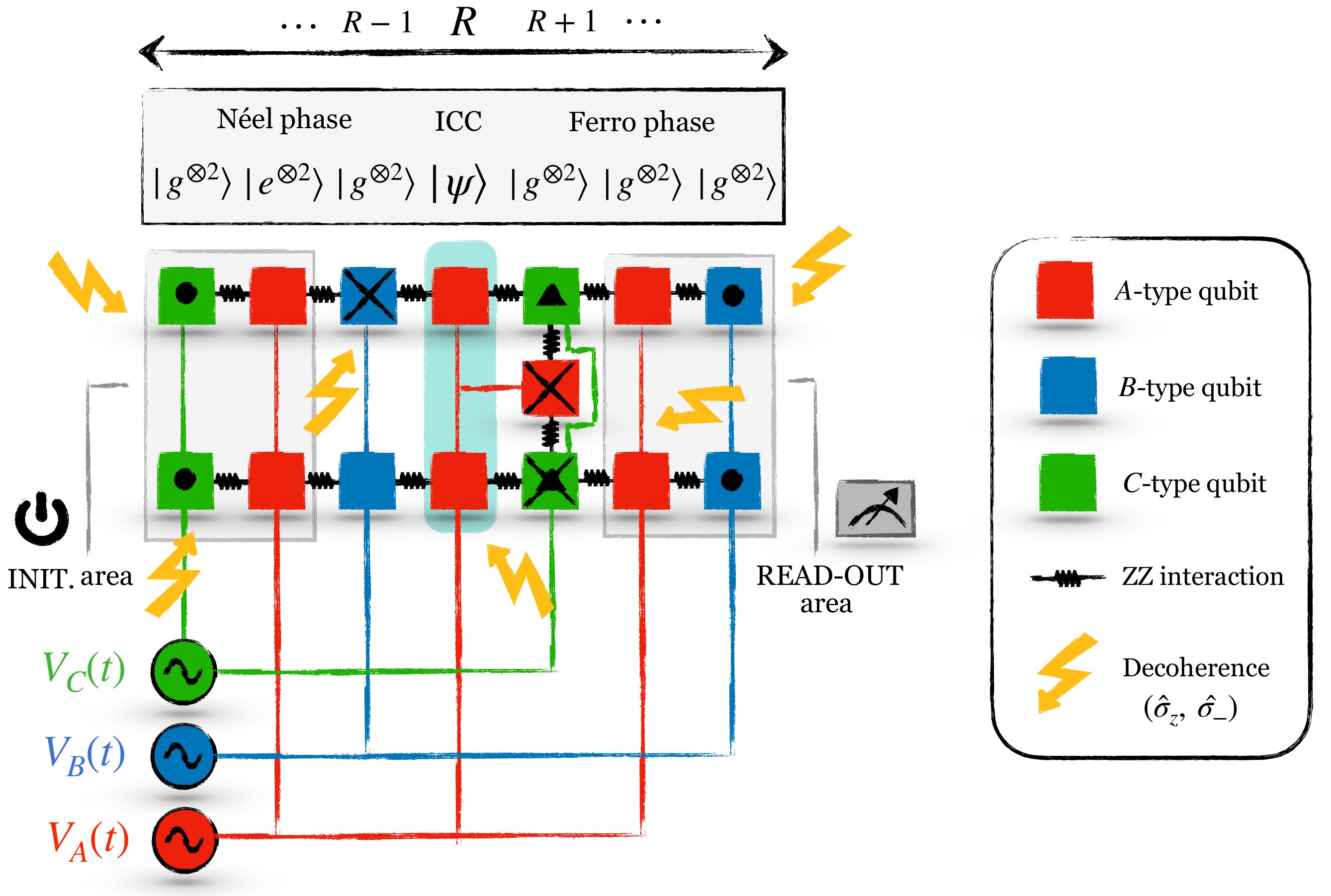}
\caption{\label{fig:ladder} Sketch of the globally-driven quantum computing ladder architecture encoding $N=2$ computational qubits. The three types of qubits $A$, $B$, and $C$, represented as red, blue, and green squares respectively, are driven by their corresponding global control classical source $V_{A,B,C}(t)$. Qubits are coupled via nearest-neighbor ZZ interactions (black springs). Crossed squares indicate qubits with doubled Rabi frequency. Black triangles and circles denote qubits with a non-standard local energy. The light-blue-highlighted column is the information carrier column (ICC), located at the interface between the N\'{e}el and polarized (ferromagnetic) phases. The first two columns constitute the initialization area and the last two form the read-out area. The yellow lightning bolts represent the presence of decoherence in the system acting on all the $2N^2+4N-1$ physical qubits.}
\end{figure}

The primary advantage of global driving is the reduction in the number of
independent control signals, which alleviates the wiring problem and limits
thermal noise at the hardware level. 
Furthermore the inherent parallelism of global control can be
advantageous for quantum error correction and fault-tolerant
operations~\cite{Bririd_2004,Kay_2005,Kay_2007,Fitzsimons_2007,
Fitzsimons_2008}.
Nevertheless, the robustness of such
architectures against dissipation---generically referred to as decoherence or
noise---remains an open and pressing question. At the experimental level,
relaxation and dephasing errors are unavoidable, and their cumulative effect
on the reliability of quantum processing units must be carefully
characterized. Such a characterization is essential not only to assess the
viability of globally-driven processors but also to design and optimize
quantum error correction schemes tailored to this specific ladder architecture.

In this work, we investigate both theoretically and numerically, using
tensor-network-based techniques, the impact of relaxation and dephasing
quantum errors on a globally-driven superconducting quantum computing
architecture where logical qubits are encoded in a subspace of the full Hilbert space~\cite{menta2024globally, cesa2023universal}. We systematically
analyze how these dissipative processes influence the spreading of quantum
information along the ladder structure and quantify their effect on the
fidelity of one- and two-qubit gate operations. Furthermore, we demonstrate
that pulse optimization constitutes an effective strategy to counteract
dissipation-induced errors and significantly extend the operational range of
the architecture.

The paper is organized as follows. In Sec.~\ref{sec:model}, we
introduce the model and the dissipative dynamics of the ladder architecture.
In Sec.~\ref{sec:operation}, we present our main numerical results on quantum
information flow and one- and two-qubit gate operations in the presence of dissipation. In Sec.~\ref{sec:mitigation}, we introduce the pulse optimization framework and demonstrate its effectiveness in mitigating decoherence. We
conclude in Sec.~\ref{sec:conclusion} with a summary of the results and an
outlook on future research directions. Technical details on the ladder
architecture, the analytical methods, and the numerical approach are provided
in the Appendices.

\section{Model and noisy dynamics}
\label{sec:model}
The system we consider here, schematically represented in
Fig.~\ref{fig:ladder},
is  based on the architecture introduced in Ref.~\cite{menta2024globally}. It consists  of three  subsets of 
qubits $A$, $B$, and $C$, arranged 
in a ladder geometry formed  by two rows  of seven qubits connected through an  additional intermediate extra qubit (the red crossed square of the figure), for a total of $N_{\rm tot}=15 $ qubits. The device is driven by   three independent  control lines, each addressing one qubit subset (solid lines in the figure). 
The
Hamiltonian  $\hat{H}(t)=\hat{H}_0 + \hat{H}_{\rm drive}(t)$    includes  a static   term $\hat{H}_0$ and 
driving contribution 
$\hat{H}_{\rm drive}(t)$. 
The former accounts for the local energies 
of the qubits and nearest-neighbor antiferromagnetic ZZ couplings $\zeta>0$ between qubits 
that belong to different subsets, 
\begin{align}
\label{H0text}
&\hat{H}_0 = \sum_{\chi \in 
\mathcal{S}}\sum_{j\in \chi} \frac{\hbar \omega_{\chi}}{2}\hat{\sigma}^{z}_{j} + \sum_{\langle j,j' \rangle} \frac{\hbar \zeta}{2}  \hat{\sigma}^{z}_{j}  \hat{\sigma}^{z}_{j'} \ , 
\end{align}
with  $\mathcal{S}:=\{A,B,C\}$, and $\sigma^{x,y,z}_j$  the  Pauli  operators of the 
qubit at site $j$. 
The driving contribution $\hat{H}_{\rm drive}(t)$ accounts instead for the action of the three classical control lines each characterized by a time-dependent Rabi frequency $\Omega_\chi(t)$, phases $\phi_\chi(t)$, and fixed carrier frequency $\omega_{\mathrm{d},\chi}$. Its explicit form, together with the non-standard local basis and the energy/Rabi shifts associated with the circle, triangle, and crossed qubits of Fig.~\ref{fig:ladder}, is reported in Appendix~\ref{sec:intro_global}. In the following, we work directly with the  effective counterpart of $\hat{H}(t)$ which governs the dynamics on the relevant timescales, as derived in Refs.~\cite{menta2024globally,menta2025building,cioni2024conveyorbelt}. This is obtained by moving into the rotating frame (RF) induced by the unitary transformation
\begin{eqnarray}
\tilde{U}(t) &=& \bigotimes_{\chi\in \mathcal{S} }
\bigotimes_{j\in \chi}\exp(i\omega_{\mathrm{d},\chi}t\,\hat{\sigma}^z_j/2),
\end{eqnarray}
associated with  the driving frequencies 
of the model, 
and applying the rotating-wave approximation (RWA), to discard the fast counter-rotating terms. 
The resulting Hamiltonian reads 
\begin{eqnarray}
\hat{H}^{\mathrm{eff}}(t) = \hat{H}_0^{\mathrm{eff}} + \hat{H}_{\mathrm{drive}}^{\mathrm{eff}}(t) \ ,
\end{eqnarray}
where $\hat{H}_0^{\mathrm{eff}}$ is obtained from (\ref{H0text}) replacing $\omega_{\chi}$
with 
$\omega_{\chi}' = \omega_{\chi} - \omega_{\mathrm{d},\chi}$, and where   
\begin{align}
\label{hdrive2-main}
&\hat{H}_{\mathrm{drive}}^{\mathrm{eff}}(t) = \sum_{\chi \in \mathcal{S}}\sum_{j\in \chi} \frac{\hbar \Omega_{\chi}(t)}{2} \Big\{\cos[\phi_{\chi}(t)] \hat{\sigma}^{x}_j + \sin[\phi_{\chi}(t)]\hat{\sigma}^{y}_j \Big\} \ .
\end{align}
Crucially, the effective drive in Eq.~\eqref{hdrive2-main} does not depend on the fast carrier frequency $\omega_{\mathrm{d},\chi}$; the latter being eliminated once using the RWA. Removing this large frequency strips away the most stringent timescale, thus permitting the tensor-network simulations of Sec.~\ref{sec:operation} to be tractable: a faithful integration of the bare model would otherwise require a time step resolving the drive period $T_{\mathrm{d},\chi} = 2\pi/\omega_{\mathrm{d},\chi} \ll T$, with $T$ denoting the total observation time. As a final remark we observe that setting 
 the resonant condition $\omega_{\mathrm{d},\chi} = \omega_{\chi} - 2\zeta$, and invoking 
 the strongly interacting regime $\eta := \zeta/|\Omega_{\chi}| \gg 1$, where the ZZ interaction dominates the drive, $\hat{H}^{\mathrm{eff}}(t)$ realizes a dynamically induced pseudo-Rydberg-blockade interaction~\cite{riccardi2026, menta2024globally}: driving on resonance selectively allows single-qubit transitions of the form $|ggg\rangle \leftrightarrow |geg\rangle$ on a site whose nearest neighbors are unexcited, while all other transitions are off-resonant and thus suppressed. This blockade is the mechanism underlying the encoding and gate operations described next.

It is now time to describe the dissipative dynamics of the ladder architecture. For this purpose  we adopt  the Gorini–Kossakowski–Sudarshan–Lindblad 
(GKSL) linear master equation (LME)~\cite{GKS, Lindblad1976} with jump operators $\hat{L}_{j,-} = \sqrt{\gamma_-}\, \hat{\sigma}^-_j$ and $\hat{L}_{j,z} = \sqrt{\gamma_z}\, \hat{\sigma}^z_j$,  associated
respectively to the energy relaxation ($T_1$) and the pure dephasing ($T_\phi$)
 of the qubit at site $j$ of the model, with on-site rates $\gamma_-$ and $\gamma_z$. 
Note that both jump operators preserve the translational invariance that is broken by the total Hamiltonian $\hat{H}(t)$. Most importantly under the RF+RWA transformation the relaxation and dephasing terms are  invariant,
leading to an effective  LME  
\begin{align}
\label{LMEeff}
\frac{\mathrm{d}}{\mathrm{d}t}\hat{\rho}(t) = &-\frac{i}{\hbar}\big[\hat{H}^{\mathrm{eff}}(t),\hat{\rho}(t)\big] \\ &+ \sum_{j,\alpha} \hat{L}_{j,\alpha} \hat{\rho}(t) \hat{L}^{\dag}_{j, \alpha} - \frac{1}{2}\sum_{j,\alpha}\big\{\hat{L}^{\dag}_{j,\alpha}\hat{L}_{j,\alpha},\hat{\rho}(t)\big\} \ , \nonumber
\end{align}
with $\alpha \in \{-, z\}$ and with $\hat{\rho}(t)$ the density matrix of the device expressed in the rotating  frame, and with the same jump operators as in the lab frame
(see Appendix~\ref{rf_jwa_LME} for details). Two assumptions implicit in this model are worth stating. The dissipator contains only $\hat{\sigma}^-$ and therefore describes pure loss at zero temperature, the relevant limit for superconducting qubits with $\hbar\omega_\chi \gg k_{\rm B} T$ and $k_{\rm B}$ being the Boltzmann constant. Its jump operators are moreover defined in the local qubit basis, whereas in the strongly interacting regime a microscopic secular derivation would yield dressed, frequency-dependent transitions: the local GKSL form adopted here is thus a standard phenomenological approximation, which we expect to capture the dominant dissipative physics of the architecture.

The dissipation rates $\gamma_-$ and $\gamma_z$ share the same angular frequency units as the Rabi frequency $\Omega_\chi$ and are therefore bare incoherent rates with dimensions of inverse time. Throughout we work with the calibration $\Omega_\chi = 10~\mu\mathrm{s}^{-1}$ and $\zeta = 200~\mu\mathrm{s}^{-1}$ (angular frequencies), which fixes $\eta = \zeta/\Omega_\chi = 20$ and sets the elementary pulse time $\pi/\Omega_\chi \simeq 314$~ns; all rates and times quoted below refer to this calibration. For an isolated qubit, the relaxation channel depletes the excited-state population at a rate $\gamma_-$, defining the relaxation time $T_1 = 1/\gamma_-$. The transverse coherence decays at the rate $1/T_2 = \gamma_-/2 + 2\gamma_z$, i.e.
\begin{equation}
\label{T2relation}
\frac{1}{T_2} = \frac{1}{2T_1} + \frac{1}{T_\phi}, \qquad \frac{1}{T_\phi} \equiv 2\gamma_z,
\end{equation}
recovering the standard relation between $T_1$, $T_2$, and the pure-dephasing time $T_\phi$. In the numerical simulations of Sec.~\ref{sec:operation}, both rates are swept over $[0,\,0.08]~\mu\mathrm{s}^{-1}$, i.e.\ up to $\gamma_\alpha/\Omega_\chi = 8\times10^{-3}$ in units of the Rabi frequency. The relaxation range $\gamma_- \in [0,\, 0.08]~\mu\mathrm{s}^{-1}$ corresponds to $T_1 \in [12.5,\, \infty)~\mu$s, with the larger rates intentionally probing the regime below the relaxation times of current superconducting platforms, which typically range from several tens to a few hundreds of microseconds.

We conclude this section by noting that a comprehensive characterization of the ladder quantum processor should also address the effects of static disorder, namely inhomogeneities in the local qubit frequencies $\omega_j$ and in the nearest-neighbor antiferromagnetic coupling strength $\zeta$. Such imperfections, which naturally arise from fabrication-induced parameter variations, constitute a major source of performance degradation, as they can hinder coherent quantum information transfer and reduce gate fidelities. A detailed analysis of these effects for the minimal globally driven ladder architecture, both with and without pulse optimization, is presented in Ref.~\cite{aiudi2026}. Since the present work is primarily concerned with the ideal architecture and the design of globally driven gate protocols, disorder-induced effects will not be considered further.

\section{Quantum information flow \& quantum gate operations}
\label{sec:operation}

In this Section, we present numerical simulations of the quantum information flow (QIF) and of one- and two-qubit gate (1QG, 2QG) operations in the presence of dephasing and relaxation dissipation.

Before turning to the numerical analysis, it is useful to recall that, in the absence of environmental noise, the ladder architecture considered here provides a universal platform for quantum computation by encoding logical information within a special subspace of the physical Hilbert space. A detailed discussion of this encoding is provided in Appendix~\ref{sec:intro_global} and in Refs.~\cite{cesa2023universal, menta2024globally, menta2025building, werner2026polynomialequivalenceglobaltransversefield}. Here we only summarize the key ingredients. In particular, the blockade condition introduced above constrains the dynamics to a structured subspace of the full $2^{N_{\rm tot}}$-dimensional Hilbert space. Following Ref.~\cite{menta2024globally}, we refer to as \emph{well-formed} those states in which each row contains a single domain wall separating a N\'eel-ordered region on the left from a polarized (ferromagnetic) region on the right. Concretely, an $N$-row well-formed state is defined as
\begin{equation}
\label{wellformed}
\ket{\mathcal{W}^{(N)}_R(\psi)} :=
\begin{pmatrix}
    \ket{\text{N\'eel}^{(1)}} \\
    \ket{\text{N\'eel}^{(2)}} \\
    \vdots \\
    \ket{\text{N\'eel}^{(N)}}
\end{pmatrix}
\otimes 
\begin{pmatrix} 
& \\
& \\
\ \ \ket{\psi_R} \\
& \\
&
\end{pmatrix}
\otimes
\begin{pmatrix}
    \ket{\text{Ferro}^{(1)}} \\
    \ket{\text{Ferro}^{(2)}} \\
    \vdots \\
    \ket{\text{Ferro}^{(N)}}
\end{pmatrix}
\end{equation}
where the information carrier column (ICC) $\ket{\psi_R}$ occupies the domain-wall interface at lattice site $R$, between the N\'eel-ordered regions $\ket{\text{N\'eel}^{(i)}} := \ket{\ldots geg}$ and the ferromagnetic regions $\ket{\text{Ferro}^{(i)}} := \ket{ggg \ldots}$ (auxiliary qubits which provide connection among the rows are also assumed to be in $|g\rangle$ state).  The ICC holds the logical information, while the two ordered regions act as a passive, classically frozen background~\cite{in-prep-transfer}. The defining property of well-formed states is that they are mapped onto one another by global pulse sequences: under the dynamical blockade constraint~\cite{riccardi2026}, the alternating drives can only shift the ICC by a single site or rotate the carrier qubit---they can never create additional domain walls. In particular, there exists a global unitary $\hat{U}_{\rm shift}$ that implements the one-site translation, \begin{equation}\hat{U}_{\rm shift} \ket{\mathcal{W}^{(N)}_R(\psi)} = \ket{\mathcal{W}^{(N)}_{R+1}(\psi)}.\end{equation} The explicit decomposition of
$\hat{U}_{\rm shift}$ operation in terms of a sequence of driving pulses, is presented in Appendix~\ref{sec:global_control}: here we only mention that
due to the presence of interactions between the rows of the systems (e.g.~the red crossed qubit element of Fig.~\ref{fig:ladder})
such decomposition does not factorize into local terms that act on the individual rows.  
The well-formed subspace is therefore closed under the available controls and constitutes the computational space of the processor.

The computation then consists of three stages. \emph{(i) Encoding.} The ICC is prepared in the initialization area (first two columns of Fig.~\ref{fig:ladder}). \emph{(ii) Processing.} Global pulses are applied selectively to the $A$-, $B$-, and $C$-type lines realizing a universal gate set: $\hat{U}_{\rm shift}$ rigidly translates the ICC through the ladder, crossed $B$/$C$ qubits implement arbitrary single-qubit rotations on the carrier they host, and a crossed $A$-type coupler enacts a two-qubit controlled-$Z$ gate between vertically adjacent logical qubits. Because distinct qubit types and crossed/regular elements respond differently to the same global field~\cite{menta2025building}, local logical operations are achieved without local addressing. \emph{(iii) Read-out.} The ICC is shifted into the read-out area (last two columns of Fig.~\ref{fig:ladder}), where the logical state is measured. All operations preserve the well-formed structure~\eqref{wellformed}, so the protocol can be viewed as the controlled propagation and manipulation of domain walls carrying quantum information.

\subsection{Quantum information flow under dissipation}
\label{sec:operation-QIF}
We begin by investigating the dissipative dynamics of the globally driven ladder architecture for $N=2$ logical qubits, focusing on the ICC translation operation. This operation constitutes the most fundamental building block of the architecture, as it underlies the coherent transport of quantum information along the ladder. We characterize it through the {global} fidelity
\begin{equation}
\label{eq:global-fidelity}
\mathcal{F} = \sqrt{\langle \Psi_{\rm target} |\hat{\rho}(T) |\Psi_{\rm target}\rangle} ,
\end{equation}
i.e.~the square root of the overlap between the full $N_{\rm tot}$-qubit states of the ladder. We adopt this ``root'' convention, which for a pure evolved state reduces to the familiar $\mathcal{F} = |\langle \Psi_{\rm target}|\Psi(T)\rangle|$; the alternative convention $\mathcal{F}^2 = \langle \Psi_{\rm target}|\hat\rho(T)|\Psi_{\rm target}\rangle$ is linear in $\hat\rho$ and is the quantity actually estimated by the trajectory average of Eq.~\eqref{eq:GRAPE_costNEW} before taking the square root. Here 
\begin{equation}
\ket{\Psi_{\rm target}} = \hat{U}_{\rm shift}\ket{\mathcal{W}_R^{(2)}(\psi_{\rm in})}=
\ket{\mathcal{W}_{R+1}^{(2)}(\psi_{\rm in})}\ ,
\end{equation} 
is the ideal noiseless target, and $\hat{\rho}(T)$ is the  evolution of  $\ket{\mathcal{W}_R^{(2)}(\psi_{\rm in})}$ obtained by integrating the LME (\ref{LMEeff}) at the end of the pulse sequence. Using the global fidelity, rather than one restricted to the carrier subsystem, ensures that any leakage of the encoded information into the surrounding lattice is fully captured. The evolved state $\hat{\rho}(T)$ is computed numerically using the quantum trajectory method combined with the MPS representation described in Appendix~\ref{sec:methods}; each data point is averaged over $N_{\rm traj}=2000$ trajectories,
\begin{equation}
\label{eq:GRAPE_costNEW}
{\mathcal{F}} \simeq \bar{\mathcal{F}}=
\left(\frac{1}{N_{\rm traj}}\sum_{k=1}^{N_{\rm traj}} \big| \langle \Psi_{\mathrm{target}} | \Psi_{\mathrm{final}}^{(k)}(T) \rangle \big|^2\right)^{1/2} \ ,
\end{equation}
and the dimensionless interaction parameter, from now on, is set to $\eta = 20$, ensuring that the system operates deep within the blockade regime.

We prepare the ICC in position $R=3$ at the interface between the N\'eel and polarized phases with both carrier qubits in the ground state, $\ket{\psi_{\rm in}} = \ket{g_1 g_2}$, where the indices $1$ and $2$ label the two logical qubits forming the ICC; the surrounding well-formed background of Eq.~\eqref{wellformed} is left implicit. This is the most robust payload the register can hold: it carries no excitation for $\hat{\sigma}^-$ to remove and no static coherence for $\hat{\sigma}^z$ to dephase. This input is therefore expected to constitute a comparatively favorable case, since it initially contains neither logical excitations susceptible to relaxation nor logical coherences susceptible to dephasing. In an interacting, globally driven system the accumulated error depends on the entire driven trajectory, so a complete characterization would require averaging over the logical input space; the values reported here should be read as representative of this favorable configuration.

\begin{figure}[t!]
\centering
\includegraphics[width=\columnwidth]{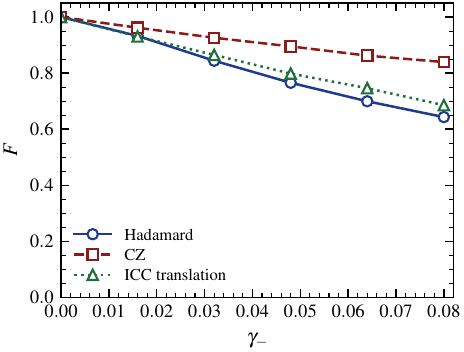}
\caption{Fidelity $\mathcal{F}$ of Eq.~(\ref{eq:global-fidelity}) for the three quantum operations considered in this work as a function of the relaxation rate $\gamma_-$ with $\gamma_z = 0$, for $N = 2$ logical qubits. Open circles (solid line) correspond to the Hadamard gate, open squares (dashed line) to the CZ gate, and open triangles (dotted line) to the ICC single-shift operation. 
The rate $\gamma_-$ is expressed in $\mu\mathrm{s}^{-1}$, with $\Omega_\chi = 10~\mu\mathrm{s}^{-1}$.}
\label{fig:fid_gamma_x} 
\end{figure}

The effect of relaxation on the ICC translation is shown by the triangular markers in Fig.~\ref{fig:fid_gamma_x}, where $\mathcal{F}$ is plotted as a function of $\gamma_-$ at $\gamma_z = 0$. The fidelity decays monotonically with increasing $\gamma_-$, from unity to $\mathcal{F} \approx 0.80$ at $\gamma_- = 0.048$ and $\mathcal{F} \approx 0.69$ at $\gamma_- = 0.08$. The mechanism is specific to the relaxation channel: because $\hat{\sigma}^-$ only de-excites, the dominant error is not the creation of spurious excitations but the depletion of the excitations that define the N\'eel order at the phase boundary. When a boundary excitation decays, the blockade condition is locally violated, the subsequent $\Pi$ pulses misfire, and the encoded information leaks and spreads across the lattice.

\begin{figure}[t!]
\centering
\includegraphics[width=\columnwidth]{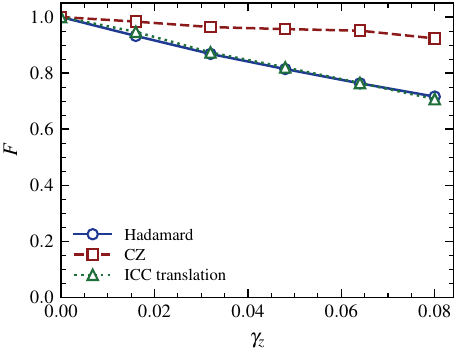}
\caption{Fidelity $\mathcal{F}$ of the same three quantum operations as in Fig.~\ref{fig:fid_gamma_x}, now as a function of the dephasing rate $\gamma_z$ with $\gamma_- = 0$, for $N = 2$ logical qubits. 
}
\label{fig:fid_gamma_z} 
\end{figure}

The complementary scenario, in which only dephasing acts on the system, is shown by the triangular markers in Fig.~\ref{fig:fid_gamma_z}. For the shift the two channels turn out to be nearly equivalent: at $\gamma_z = 0.048$ the fidelity is $\mathcal{F} \approx 0.82$, close to the $\mathcal{F} \approx 0.80$ reached under relaxation at the same rate in Fig.~\ref{fig:fid_gamma_x}, and at $\gamma_z = 0.08$ the two read $\mathcal{F} \approx 0.71$ and $\mathcal{F} \approx 0.69$, respectively. The reason is that, with the carriers in $\ket{g}$, the logical payload is left untouched by both channels, so the whole fidelity loss comes from the ordered background. Dephasing only scrambles the relative phases of that background, whereas relaxation removes part of its excitations and perturbs the blockade pattern directly; this is why relaxation stays the slightly more harmful of the two even here.

Relaxation is the more damaging channel for every operation studied (e.g.~for the Hadamard, $\mathcal{F}\approx0.64$ under $\gamma_-=0.08$ versus $\mathcal{F}\approx 0.72$ under $\gamma_z=0.08$). This asymmetry is the more remarkable because equal numerical values $\gamma_- = \gamma_z$ do not correspond to equally strong channels for an isolated qubit: from Eq.~\eqref{T2relation} the dephasing contribution to the transverse decay, $2\gamma_z$, is four times the relaxation contribution, $\gamma_-/2$. Relaxation nevertheless dominates the fidelity loss, confirming that the damage is mediated by the blockade mechanism rather than by bare coherence decay. This reveals that the dominant source of infidelity is the disruption of the blockade mechanism by the loss of population-encoded information, to which $\hat{\sigma}^-$ couples directly.

\subsection{One-qubit gate}
We now turn to single-qubit gate operations in the presence of dissipation, focusing on the Hadamard gate implemented via the pulse sequence described in Appendix~\ref{sec:global_control}, which exploits the differential evolution of crossed versus regular qubits. During the gate, the target computational qubit located on the crossed element of the ICC undergoes a rotation $\hat{\mathbb{R}}(\theta, \vec{n})$, while all other qubits in the ladder ideally remain stationary. The fidelity is computed following the same protocol as in Eq.~(\ref{eq:global-fidelity}), where  $\ket{\Psi_{\rm target}}$ and $\hat{\rho}(T)$ are now the ideal and noisy Hadamard-transformed counterpart of the well-formed state with $|\psi_{\rm in}\rangle =|g_1g_2\rangle$ at position $R=3$ of the device of Fig.~\ref{fig:ladder},  and each data point is averaged over $N_{\mathrm{traj}} = 2000$ quantum trajectories.

The results for the Hadamard gate under relaxation dissipation are shown by the circular markers in Fig.~\ref{fig:fid_gamma_x}. The Hadamard decays \emph{faster} than the ICC translation, despite its shorter pulse sequence: at $\gamma_- = 0.048$ the Hadamard fidelity is $\mathcal{F} \approx 0.77$ against $\mathcal{F} \approx 0.80$ for the shift, and at $\gamma_- = 0.08$ they reach $\mathcal{F} \approx 0.64$ and $\mathcal{F} \approx 0.69$, respectively, all three operations starting from the same initial state. Concretely, the Hadamard gate duration is shorter than that of the ICC translation, so the faster decay cannot be attributed to gate duration alone.
The relaxation-over-dephasing channel asymmetry persists for the Hadamard, as shown by the circular markers in Fig.~\ref{fig:fid_gamma_z}: at $\gamma_z = 0.048$ the Hadamard fidelity stands at $\mathcal{F} \approx 0.81$, and at $\gamma_z = 0.08$ at $\mathcal{F} \approx 0.72$. The decay is again slower than under relaxation dissipation, consistent with the channel asymmetry established in Sec.~\ref{sec:operation-QIF}.

Relaxation remains the main source of infidelity for the Hadamard gate as well. The underlying physical mechanism mirrors the ICC translation case: relaxation errors affect not only the target computational qubit but also all spectator qubits coupled through the ladder, including the ancillary qubits that maintain the Néel and polarized phases. Since the blockade mechanism relies on the correct population of neighboring qubits to selectively enable or suppress transitions, a relaxation on any spectator qubit can disrupt the controlled evolution of the target qubit. This collective sensitivity to relaxation noise is a distinctive feature of the globally driven architecture, where every physical qubit participates in the gate operation through the blockade constraint, even when only a single computational qubit is being rotated.

\subsection{Two-qubit gate}

The controlled-Z (CZ) gate provides the entangling operation required for universal quantum computation in the ladder architecture. It is implemented through a pulse sequence acting on the $A$-type qubits, as described in Appendix~\ref{sec:global_control}, with the ICC positioned at a column containing a crossed $A$-type coupler between two adjacent logical qubits. The gate exploits the blockade mechanism: a $2\pi$ rotation on the crossed coupler acquires a conditional phase that depends on the state of both neighboring computational qubits, thereby realizing the entangling operation $\hat{U}_{A,2} = \hat{Z}_{A^\times}$.

The numerical investigation of the CZ gate under relaxation and dephasing dissipation follows the same protocol as for the previous operations. In this case the fidelity \eqref{eq:global-fidelity} computed starting from the well-formed input state with $|\psi_{\rm in}\rangle =|g_1g_2\rangle$ at position $R=5$, is averaged over $N_{\mathrm{traj}} = 2000$ quantum trajectories at interaction parameter $\eta = 20$, and the results are shown by the square markers in Figs.~\ref{fig:fid_gamma_x} and~\ref{fig:fid_gamma_z}.

The CZ pulse sequence is the least affected by dissipation among the three operations considered, under both relaxation and dephasing dissipation. Under relaxation dissipation, the fidelity remains above $\mathcal{F} \approx 0.84$ across the full $\gamma_-$ range of Fig.~\ref{fig:fid_gamma_x}, well above the values reached by the ICC translation ($\mathcal{F} \approx 0.69$) and the Hadamard gate ($\mathcal{F} \approx 0.64$) at $\gamma_- = 0.08$. Under dephasing dissipation, the CZ fidelity remains above $\mathcal{F} \approx 0.95$ at $\gamma_z = 0.048$ and above $\mathcal{F} \approx 0.92$ at $\gamma_z = 0.08$, decaying at a markedly slower rate than the other two operations across the entire $\gamma_z$ window.

This robustness is a direct consequence of the gate duration. The CZ operation reduces to a single $2\pi$ rotation on the crossed coupler $A^\times$, with total duration $T_{\rm CZ} \sim \pi/\Omega_{\chi}$, roughly an order of magnitude shorter than both the multi-step ICC translation sequence and the Hadamard pulse sequence. Since the dissipative contribution to the time-evolved density matrix accumulates over the gate duration, a shorter gate inherits correspondingly less decoherence, regardless of the dissipation channel. The CZ gate thus benefits from a built-in advantage simply by virtue of being fast. This observation already foreshadows the central message of the next Section: shortening the gate duration through pulse optimization is an effective strategy to mitigate the impact of dissipation, even when the underlying dissipation rates cannot themselves be reduced.

\section{Mitigating decoherence: pulse optimization}
\label{sec:mitigation}

The numerical results of Sec.~\ref{sec:operation} establish that both dissipation channels degrade the fidelity of quantum operations in the globally driven ladder architecture, with the relaxation channel emerging as the dominant source of infidelity for all three operations considered. A natural question is then whether operational strategies exist to mitigate this degradation without modifying the hardware. In this Section, we argue that pulse optimization, by producing significantly shorter gate sequences, constitutes an effective approach to reducing the impact of decoherence. We focus on the Hadamard gate under relaxation dissipation, which combines the most damaging noise channel identified in Sec.~\ref{sec:operation} with a gate duration that leaves substantial room for improvement. The two-qubit CZ gate is not considered here, as its inherently short pulse sequence leaves little room for further reduction over the parameter range explored in this work.

\subsection{Gate time and decoherence accumulation}

The key observation underlying our mitigation strategy is that decoherence accumulates throughout the gate execution. For a quantum gate of duration $T_{\rm gate}$ subject to Markovian noise, the fidelity degrades monotonically with the exposure time to the dissipative environment---a direct consequence of the Lindblad master equation, in which the dissipative contribution to the time-evolved density matrix grows with the gate duration. The numerical results of Sec.~\ref{sec:operation} confirm this picture: for both the ICC translation and the Hadamard gate, the fidelity decreases monotonically with increasing dissipation rates at fixed gate duration. The immediate implication is that shortening the gate time limits the total decoherence accumulated and thereby improves the achievable fidelity, provided the optimized pulse sequence faithfully implements the target unitary without introducing additional errors.

\subsection{Pulse optimization via MPS-based \texttt{GRAPE}}
\label{sec:mps_GRAPE}

To exploit this observation, we have developed a pulse optimization engine implementing the gradient ascent pulse engineering (\texttt{GRAPE}) methodology~\cite{Khaneja2005_GRAPE}, based on MPS-based time evolution combined with the Adam optimizer~\cite{adam2017}. Gradients of the cost function are computed via forward and backward propagation of quantum states using the TDVP integrator. The optimization maximizes a noiseless cost of the same form as Eq.~\eqref{eq:GRAPE_costNEW} by averaging over a training set of $N_{\rm train}$ initial- and target-state pairs.

The control amplitudes are bounded within $[-1, 1]$ and discretized with time steps of approximately $1$ nanosecond, compatible with state-of-the-art experimental hardware. The key advantage of working within the MPS framework is the ability to optimize pulses directly on the full $N_{\rm tot} = 15$-qubit ladder architecture rather than on reduced subsystems.

In Ref.~\cite{aiudi2026}, we demonstrated that \texttt{GRAPE}-optimized pulse sequences can recover high gate fidelities in the presence of realistic fabrication-induced disorder while operating at a fraction of the gate time required by standard rectangular pulse protocols. These results motivate us to leverage pulse optimization not only to overcome the unavoidable fabrication-induced disorder present in any realistic experimental device but also, through the significantly shorter gate times that optimized pulses achieve, to reduce the exposure to dynamical decoherence.

\subsection{Decoherence mitigation for the one-qubit gate}
\label{sec:mitigation_1q}

We now put the above argument to a direct numerical test on the single-qubit Hadamard gate in the presence of relaxation dissipation. We compare the fidelity obtained with the standard rectangular pulse protocol described in Appendix~\ref{sec:global_control} to that achieved with a \texttt{GRAPE}-optimized pulse sequence generated within the framework of Sec.~\ref{sec:mps_GRAPE}. Both protocols implement the same target unitary---the Hadamard gate acting on the computational qubit located on the crossed element of the ICC---but differ substantially in their total duration: the standard protocol requires $T_{\rm std} \approx 3000$ ns, while the optimized one completes the same operation in $T_{\rm opt} \approx 320$ ns, nearly an order of magnitude shorter. To reflect a realistic experimental setting, the ladder is simulated with $2\%$ relative disorder on the local qubit frequencies, mimicking fabrication-induced imperfections. In both cases, the fidelity is evaluated as in Eq.~\eqref{eq:GRAPE_costNEW} with $N_{\mathrm{traj}} = 2000$ quantum trajectories and interaction parameter $\eta = 20$. The dissipation rates $\gamma_-$ and $\gamma_z$ are expressed in the same angular frequency units as the Rabi frequency $\Omega_\chi$, so that the associated decoherence time is $1/\gamma_\alpha$.

Before turning to the dissipative regime, we verify that the optimized pulse implements the target rotation over a one-parameter family of input states spanning the full population range, and not merely the specific training states. To this end, we evaluate its fidelity in the dissipation-free limit $\gamma_- = \gamma_z = 0$ while varying the initial state of the target qubit as $\ket{\psi_{\mathrm{initial}}(p)} = \big(\sqrt{1-p}\ket{g_1} + \sqrt{p}\ket{e_1}\big)\ket{g_2}$ with $p \in [0,1]$, following the parametrization introduced for the ICC translation.

\begin{figure}[t]
\centering
\includegraphics[width=\columnwidth]{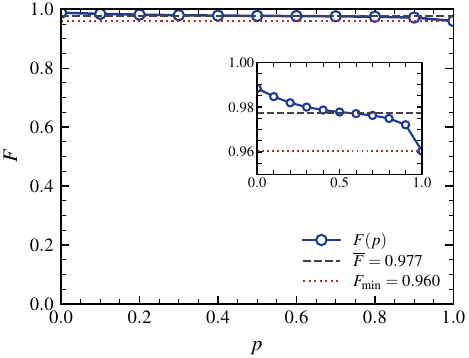}
\caption{Fidelity $\mathcal{F}$ of the \texttt{GRAPE}-optimized Hadamard gate in the dissipation-free limit $\gamma_- = \gamma_z = 0$, as a function of the initial state parameter $p$ of the target computational qubit with $\phi=0$ for $N = 2$ logical qubits. The optimized pulse has duration $T_{\rm opt} \approx 320$ ns and the ladder is simulated with $2\%$ relative disorder on the local qubit frequencies. The solid curve shows $\mathcal{F}(p)$, while the dashed and dotted lines mark the average $\overline{\mathcal{F}} \approx 0.977$ and the minimum $\mathcal{F}_{\mathrm{min}} \approx 0.960$ over the full range of $p$ (the overbar here denotes the average over $p$, not the trajectory average of Eq.~\eqref{eq:GRAPE_costNEW}). The inset zooms into the narrow fidelity window, where the slow decline toward $p = 1$ becomes visible. The fidelity remains essentially flat across all $p$, confirming that the optimized pulse acts as the intended Hadamard rotation for all sampled population at $\phi = 0$.
}
\label{fig:hadamard_opt_p} 
\end{figure}

\begin{figure}[t]
\centering
\includegraphics[width=\columnwidth]{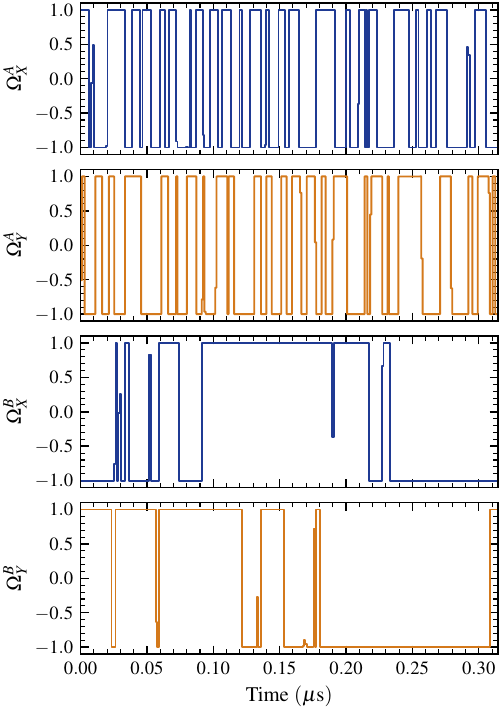}
\caption{\texttt{GRAPE}-optimized control pulses implementing the Hadamard gate on the target computational qubit for $N = 2$ logical qubits. The top panels display the in-phase ($\Omega_A^X$) and quadrature ($\Omega_A^Y$) components of the drive applied to the $A$-type qubits, while the bottom panels show the corresponding components ($\Omega_B^X$, $\Omega_B^Y$) for the $B$-type qubits. The control amplitudes are normalized to the Rabi frequency $\Omega_\chi$ and bounded within $[-1, 1]$, with a time discretization on the order of nanoseconds, compatible with state-of-the-art experimental hardware. The total gate duration is $T_{\rm opt} \approx 320$ ns. Drives on the $C$-type qubits are not shown as they remain inactive for this single-qubit gate operation.}
\label{fig:hadamard_grape_pulses} 
\end{figure}

The resulting fidelity is shown in Fig.~\ref{fig:hadamard_opt_p}, where it remains essentially flat over the entire range of $p$, with an average over $p$ of ${\mathcal{F}}_{\rm ave} \approx 0.98$ and a minimum of $\mathcal{F}_{\rm min} \approx 0.96$. This confirms that the optimized pulse implements the intended rotation over the whole sampled family which spans the populations from $\ket{g}$ to $\ket{e}$ at fixed relative phase $\phi = 0$, so that phase-sensitive errors are not probed and provides a meaningful baseline against which the impact of dissipation can be assessed. The corresponding optimized control sequence is shown in Fig.~\ref{fig:hadamard_grape_pulses}, where the in-phase and quadrature components of the drives applied to the $A$- and $B$-type qubits are displayed over the full gate duration.

\begin{figure}[t]
\centering
\includegraphics[width=\columnwidth]{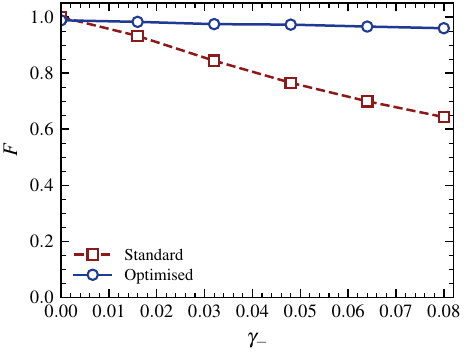}
\caption{Hadamard gate fidelity $\mathcal{F}$ as a function of the relaxation rate $\gamma_-$ with $\gamma_z = 0$ for $N = 2$ logical qubits. The red dashed curve corresponds to the standard rectangular pulse protocol with gate duration $T_{\rm std} \approx 3000$ ns, while the blue solid curve corresponds to the \texttt{GRAPE}-optimized pulse with $2\%$ relative disorder on the local qubit frequencies and gate duration $T_{\rm opt} \approx 320$ ns. The rate $\gamma_-$ is expressed in the same angular frequency units as the Rabi frequency $\Omega_\chi$. 
}
\label{fig:hadamard_opt_vs_std} 
\end{figure}

Having established that the optimized pulse faithfully implements the Hadamard gate in the absence of dissipation, we now assess its performance as the relaxation channel is switched on. The results are shown in Fig.~\ref{fig:hadamard_opt_vs_std}, where the fidelity is plotted as a function of $\gamma_- \in [0, 0.08]$ for both protocols, with $\gamma_z = 0$. In the dissipation-free limit, the standard pulse recovers essentially unit fidelity, while the optimized pulse sits at the baseline of $\mathcal{F} \approx 0.98$ set by the disordered ensemble, consistent with Fig.~\ref{fig:hadamard_opt_p}. As soon as relaxation dissipation is switched on, however, the two protocols diverge markedly. The standard pulse fidelity drops steadily, reaching $\mathcal{F} \approx 0.77$ at $\gamma_- = 0.048$ and $\mathcal{F} \approx 0.64$ at $\gamma_- = 0.08$. In contrast, the optimized pulse stays above $\mathcal{F} \approx 0.96$ across the entire range, declining only from its $\mathcal{F}\approx0.98$ disorder baseline to $\mathcal{F}\approx0.96$ at $\gamma_- = 0.08$.

This pronounced resilience is fully consistent with the physical picture developed in the preceding subsections. Since both protocols realize the same target unitary on the same architecture, the important difference is the time spent evolving under the dissipative dynamics. The nearly tenfold reduction in gate duration translates directly into a correspondingly reduced exposure to the relaxation channel, which in turn preserves the integrity of the blockade mechanism on which the single-qubit gate relies.
Crucially, this gain is obtained without any modification of the hardware or of the noise model: it stems entirely from the shape of the control pulse, a degree of freedom that can in principle be exploited on any realistic experimental platform.

These findings demonstrate that pulse optimization constitutes a practical and effective decoherence mitigation strategy at the single-qubit level in the globally driven ladder architecture. Table~\ref{tab:summary} collects the fidelities of all the operations and protocols discussed so far, and quantifies the improvement delivered by the optimized pulses.

\begin{table}[t]
\begin{ruledtabular}
\begin{tabular}{lcccc}
\multicolumn{5}{c}{(a) Operational primitives, standard pulses}\\[2pt]
\hline
 & \multicolumn{2}{c}{Relaxation} & \multicolumn{2}{c}{Dephasing}\\
Operation & $\gamma_-\!=\!0.048$ & $0.08$ & $\gamma_z\!=\!0.048$ & $0.08$\\
\hline
ICC translation & $0.80$ & $0.69$ & $0.82$ & $0.71$\\
Hadamard (1QG)  & $0.77$ & $0.64$ & $0.81$ & $0.72$\\
CZ (2QG)        & $0.90$ & $0.84$ & $0.96$ & $0.92$\\[2pt]
\hline
\multicolumn{5}{c}{(b) Hadamard gate under relaxation: mitigation}\\[2pt]
\hline
Protocol & $T_{\rm gate}$ & $\mathcal{F}(0.048)$ & $\mathcal{F}(0.08)$ & $1-\mathcal{F}(0.08)$\\
\hline
Standard        & $3000$~ns & $0.77$ & $0.64$ & $0.36$\\
\texttt{GRAPE}  & $320$~ns  & $0.97$ & $0.96$ & $0.04$\\
\hline
Gain & $9.4\times$ & --- & --- & $9\times$\\
\end{tabular}
\end{ruledtabular}
\caption{\label{tab:summary} Summary of the global fidelity $\mathcal{F}$ of Eq.~\eqref{eq:global-fidelity} for $N=2$ logical qubits, $\eta=20$ and $N_{\rm traj}=2000$; all rates are in $\mu\mathrm{s}^{-1}$. Panel~(a): the three operational primitives driven by the standard rectangular pulses of Appendix~\ref{sec:global_control}, under pure relaxation ($\gamma_z=0$) and pure dephasing ($\gamma_-=0$). Panel~(b): the Hadamard gate under relaxation for the standard and the \texttt{GRAPE}-optimized protocols of Sec.~\ref{sec:mitigation}, together with the residual infidelity $1-\mathcal{F}$ at the largest rate explored. The nearly tenfold compression of the gate duration is accompanied by a comparable reduction of the residual infidelity. Note that the optimized entries are obtained \emph{including} $2\%$ static disorder on the local qubit frequencies, which is absent from the standard protocol and already caps the optimized fidelity at $\overline{\mathcal{F}}\approx0.977$ for $\gamma_-=\gamma_z=0$; the comparison is therefore conservative.}
\end{table}

\section{Conclusion}
\label{sec:conclusion}

In this work, we investigated the dissipative dynamics of a 
globally driven superconducting quantum computing architecture based on a 
quasi-two-dimensional ladder geometry. Using tensor-network methods combined 
with the quantum trajectory approach, we have systematically characterized 
the impact of two fundamental decoherence channels---relaxation and dephasing 
processes---on three key operational primitives of the ladder architecture: 
quantum information flow via the ICC single-shift operation, single-qubit gate 
operations exemplified by the Hadamard gate, and two-qubit entangling 
operations realized through the CZ gate.

Our analysis identified relaxation as the more harmful of the two channels
for all three operations studied, with the Hadamard gate being the worst
case scenario: the fidelity falls to $\mathcal{F} \approx 0.64$ at
$\gamma_- = 0.08$, against $\mathcal{F} \approx 0.72$ under dephasing at
the same rate. The origin of this sensitivity is the structure of the
global control protocol: relaxation depletes the excitations that encode
the ICC at the phase boundary, the blockade condition is locally violated,
and the encoded information leaks and spreads incoherently across the
lattice. The noise sensitivity of the ladder architecture is therefore
shaped by the multi-step pulse protocol, not by the bare rates of the
individual dissipation channels alone. This worst case, the Hadamard gate
under relaxation, is the one we selected for the mitigation study of
Sec.~\ref{sec:mitigation}.

Among the three operations studied, the two-qubit CZ pulse sequence proves the least affected by both dissipation channels. This robustness has a simple structural origin: the CZ operation  reduces to a single $2\pi$ rotation of the crossed $A$-type coupler mediated  by the blockade interaction between adjacent computational qubits, with a  total gate time $T_{\rm CZ} \sim \pi/\Omega_\chi$ roughly an order of  magnitude shorter than the multi-step ICC translation and Hadamard sequences.  The resulting reduced exposure to both relaxation and dephasing processes  directly translates into higher fidelities across the full dissipation range  explored, confirming that gate duration is the primary lever controlling  decoherence accumulation in this architecture.

These considerations motivate the use of pulse optimization as a hardware-free  mitigation strategy. Implementing \texttt{GRAPE}-based pulse optimization  within the MPS framework, we generated pulse sequences whose total gate  duration is nearly an order of magnitude shorter than the standard rectangular  pulse protocols: $T_{\rm opt} \approx 320$~ns versus $T_{\rm std} \approx  3000$~ns for the Hadamard gate, corresponding to a time-overhead reduction  factor of $T_{\rm std}/T_{\rm opt} \approx 9.4$. This dramatic compression of  the gate time is achieved while simultaneously absorbing the effect of $2\%$  fabrication-induced disorder on the local qubit frequencies, confirming that  both static and dynamical sources of error can be addressed within the same  optimization framework. The optimized Hadamard pulse stays above $\mathcal{F} \approx 0.96$ across the whole range of relaxation rates considered, compared to $\mathcal{F} \approx 0.64$ at $\gamma_- = 0.08$ for the standard protocol---a gain that requires no change to the hardware or the noise model, only a reshaping of the control pulse.

Taken together, these results establish a coherent physical picture of  decoherence in globally driven quantum processors and provide actionable  guidelines for their experimental implementation. The dominant vulnerability  of the architecture lies in its sensitivity to relaxation errors acting on the  spectator qubits that maintain the Néel and polarized phases, and the most  effective mitigation strategy consists in minimizing the gate duration through  pulse optimization rather than through hardware modifications. The fact that  this strategy simultaneously addresses both dynamical decoherence and static  fabrication disorder makes it particularly attractive for near-term  experimental realizations, and connects naturally to broader developments in  quantum optimal control beyond \texttt{GRAPE}~\cite{hu2025universal,gargiulo2026obstructions,Bondar2025}.

Two natural extensions of the present work suggest themselves. First, we have  assumed a purely Markovian noise model; incorporating non-Markovian effects~\cite{Pellitteri2026},  thermal noise, shot noise, and residual ZZ cross-talk would bring the model  closer to realistic experimental conditions. Second, the MPS-\texttt{GRAPE}  framework developed here provides a natural starting point for designing  fault-tolerant pulse sequences~\cite{Bririd_2004,Kay_2005,Kay_2007,Fitzsimons_2007,Fitzsimons_2008}  specifically tailored to the ladder architecture, where the global drive  constraint imposes stringent requirements on the structure of correctable  error syndromes.

\section*{Data availability}
The data reproducing all numerical results are available at~\cite{Data}.

\acknowledgments
We thank R.~Aiudi and F.~Cioni for useful discussions in the early stages of the project.
We gratefully acknowledge computational resources of the Center for High Performance Computing (CHPC) at SNS. At the time of their contributions, authors affiliated with Planckian are either employees at Planckian or research collaborators with Planckian.

\bibliography{biblio}

\appendix

\section{The ladder architecture}
\label{sec:intro_global}

We summarize the physical properties of the globally driven ladder architecture proposed in Ref.~\cite{menta2024globally}, and provide the explicit Hamiltonian and encoding details referenced in the main text.

The architecture is a ladder of short-range interacting qubits of three types, $\mathcal{S}:=\{A,B,C\}$, arranged along the ladder axis in the periodic pattern $\ldots CABACABA\ldots$ and isotropic in the transverse direction (Fig.~\ref{fig:ladder}). For $N$ rows the lattice has $2N+3$ columns---the three extra columns hosting the information carrier column (ICC), the initialization area, and the read-out area---while inter-row coupling is mediated by $N-1$ coupler qubits, for a total of $N_{\rm tot} = 2N^2 + 4N - 1$ physical qubits. Throughout we use the $N=2$ instance ($N_{\rm tot}=15$), the minimal fully functional processor and the optimal choice for numerical study. Each qubit is coupled to its intra-row nearest neighbors through a nearest-neighbor antiferromagnetic ZZ interaction of strength $\zeta$, and selected $B$- and $C$-type qubits are additionally ZZ-coupled to the $A$-type couplers bridging vertically adjacent rows.

Qubits marked with a circle ($\bullet$) or a triangle ($\blacktriangle$) carry shifted local energies $\omega_\chi^{\bullet}:=\omega_\chi-\zeta$ and $\omega_\chi^{\blacktriangle}:=\omega_\chi+\zeta$, reflecting their $1$ and $3$ nearest-neighbor connections (a regular qubit has $2$). Crossed qubits (crossed squares) are driven at twice the Rabi frequency of their regular counterparts, $\Omega_{\chi^\times}=2\Omega_{\chi^{\rm r}}$; accordingly each qubit is labeled both by its type $\chi$ and by whether it is regular or crossed, $\chi\in\{\chi^{\rm r},\chi^\times\}$. The local basis is $\mathcal{B}_j = \{\vert g_j\rangle, \vert e_j \rangle\}$, with $\vert g_j \rangle := (0,1)^{\intercal}$ and $\vert e_j \rangle := (1,0)^\intercal$.

The total Hamiltonian decomposes as $\hat{H}(t) = \hat{H}_0 + \hat{H}_{\rm drive}(t)$. Its static part $\hat{H}_0$ is given in Eq.~\eqref{H0text} of the main text, while the driving contribution, engineered via the three classical control lines, reads
\begin{align}
\label{Hdrive}
\hat{H}_{\rm drive}(t) = \sum_{\chi \in \mathcal{S}}\sum_{j \in \chi}\hbar\Omega_\chi(t) \sin[\omega_{\mathrm{d},\chi}t + \phi_{\chi}(t)] \hat{\sigma}^{y}_j \ ,
\end{align}
where $j \in \chi$ is an integer lattice site index, $\omega_{\mathrm{d},\chi} > 0$ is the drive frequency of the $\chi$-type pulse, and $\Omega_{\chi}(t),\,\phi_{\chi}(t) \in \mathbb{R}$ are the time-dependent Rabi frequency and global phase. Crossed qubits are driven at twice the Rabi frequency, $\Omega_{\chi}(t) \to 2\Omega_\chi(t)$.

Moving to the frame rotating at the drive frequencies, generated by the unitary $\tilde{U}(t)$ introduced in Sec.~\ref{sec:model}, and invoking the rotating-wave approximation (RWA) with the resonance condition $\omega_{\mathrm{d},\chi} = \omega_{\chi} - 2\zeta$, the total Hamiltonian becomes $\hat{H}^{\mathrm{eff}}(t) = \hat{H}_0^{\mathrm{eff}} + \hat{H}_{\mathrm{drive}}^{\mathrm{eff}}(t)$, with
\begin{align}
\label{Hfinal}
\hat{H}_0^{\mathrm{eff}} = \sum_{\chi \in \mathcal{S}}\sum_{j \in \chi} \frac{\hbar\omega_{\chi}'}{2}\hat{\sigma}^{z}_{j} + \sum_{\langle j,j' \rangle} \frac{\hbar \zeta}{2}  \hat{\sigma}^{z}_{j}  \hat{\sigma}^{z}_{j'} \ ,
\end{align}
where $\omega_{\chi}' = \omega_{\chi} - \omega_{\mathrm{d},\chi}$. The frame transformation alone leaves the drive in the intermediate form
\begin{align}
\label{hdrive1}
\hat{H}_{\mathrm{drive}}^{\mathrm{eff}\,\prime}(t) = \sum_{\chi \in \mathcal{S}}\sum_{j \in \chi} \hbar V_{\chi}(t)\big[\cos(\omega_{\mathrm{d},\chi}t) \hat{\sigma}^y_j + \sin(\omega_{\mathrm{d},\chi}t) \hat{\sigma}^x_j\big] \ ,
\end{align}
with $V_{\chi}(t) = \Omega_\chi(t) \sin[\omega_{\mathrm{d},\chi}t + \phi_{\chi}(t)]$, the prime distinguishing it from its post-RWA counterpart. Applying the RWA once more eliminates the fast carrier $\omega_{\mathrm{d},\chi}$ and reduces the drive to the slowly-varying $\hat{H}_{\mathrm{drive}}^{\mathrm{eff}}(t)$ of Eq.~\eqref{hdrive2-main} of the main text.

A logical qubit is stored as a domain wall at the interface between a N\'eel-ordered and a polarized (ferromagnetic) region of a row, with the carrier qubit sitting at the boundary~\cite{cesa2023universal, menta2024globally}. Denoting by $j=(r,R)$ the physical site sitting in row $r$ and column $R$, the carrier hosted there reads $\ket{\psi_R^{(r)}} = \alpha_r\ket{g_j} + \beta_r\ket{e_j}$ in the local basis $\mathcal{B}_j$, with $|\alpha_r|^2+|\beta_r|^2=1$. The $N$ carriers sharing the column $R$ constitute the ICC $\ket{\psi_R} = \bigotimes_{r=1}^{N}\ket{\psi_R^{(r)}}$ entering the well-formed state of Eq.~\eqref{wellformed}, which the global pulse sequences transport coherently from one column to the next through the elementary shift $\hat{U}_{\rm shift}$, whose explicit pulse decomposition is given in Eq.~\eqref{equshiftrev} below. Universal control is provided by the crossed qubits, which act as localized control elements despite the uniform drive: the $N$ crossed $B$/$C$ qubits implement arbitrary single-qubit rotations on the carrier in their row, while the $N-1$ crossed $A$-type couplers realize controlled-$Z$ gates between vertically adjacent logical qubits~\cite{menta2025building, menta2024globally, cioni2024conveyorbelt, cesa2023universal}. The explicit pulse sequences realizing these operations are derived in Appendix~\ref{sec:global_control}.

\section{Global control for quantum information flow}
\label{sec:global_control}

\subsection{Universal global control}

The real-time evolution of the ladder architecture is governed by the time-dependent drive Hamiltonian $\hat{H}^{\mathrm{eff}}_{\mathrm{drive}}(t)$. The total observation time is denoted $T$ and defined as $T = \max(\tau_{\mathrm{tot}})$, where $\tau_{\rm tot}$ is the total time window. The latter is partitioned into several sub-windows; during each window $\tau_{\chi}$, the control signal $V_{\chi}(t)$ is active and held constant while all others are set to zero. This implies that the pair $(\Omega_{\chi}(t), \phi_{\chi}(t))$ is constant within each window, so that $\hat{H}_{\rm drive}^{\rm eff}(t)$, defined in Eq.~\eqref{hdrive2-main}, becomes time-independent within each interval, greatly simplifying the numerical calculations. We impose the following disjoint activation conditions:
\begin{align}
& V_A(t) \neq 0 \Longrightarrow V_B(t)=V_C(t)=0, \\
& V_B(t) \neq 0 \text{ or } V_C(t) \neq 0 \Longrightarrow V_A(t) = 0.
\end{align}
The global control thus enforces selective activation of different qubit groups, avoiding interference between them. This allows the total evolution to be decomposed into a union of disjoint time windows, $\tau_{\rm tot} = \bigcup_{\ell} \tau_{\chi_{\ell}}$ with $\chi_{\ell} \in \mathcal{S}$, within each of which the drive Hamiltonian can be treated as time-independent: $\hat{H}_{\rm drive}^{\rm eff}(t) = \hat{H}_{\rm drive,\ell}^{\rm eff}$ for all $t \in \tau_{\chi_{\ell}}$. The time-ordered unitary evolution over the full time window then simplifies to:
\begin{equation}
\label{stringSUP}
\hat{U}_{\rm tot} = \mathcal{T} \exp\left[ -\frac{i}{\hbar} \int_{0}^{T} \mathrm{d}t'\, \hat{H}_{\rm drive}^{\rm eff}(t')\right] = \hat{U}_{\chi_{\ell}} \cdots \hat{U}_{\chi_2}\hat{U}_{\chi_1},
\end{equation}
where $\mathcal{T}$ is the time-ordering operator and $\hat{U}_{\chi_{\ell}} = \exp\big(-i \hat{H}^{\rm eff}_{\rm drive, \ell} \tau_{\chi_\ell} / \hbar \big)$ is the unitary operator associated with the global drive Hamiltonian during the $\ell$-th interval of duration $\tau_{\chi_\ell}$.

In the strongly interacting regime $\eta \gg 1$, transitions (excitations or de-excitations) induced by $\hat{H}^{\rm eff}_{\rm drive}$ are suppressed whenever the time-evolved many-body state has at least one nearest neighbor of the driven $\chi$-type qubit in the excited state. In this limit, each $\hat{U}_{\chi_\ell}$ in Eq.~\eqref{stringSUP} can be expressed in terms of controlled unitary gates~\cite{menta2024globally} as:
\begin{equation}
\label{eq:Uell}
\hat{U}_{\chi_\ell} = \prod_{j \in \chi_{\ell}} \Big(\mathds{1}_j \otimes \hat{Q}_{\langle j \rangle} + \hat{\mathbb{U}}_{\chi_\ell,j} \otimes \hat{P}_{\langle j \rangle} \Big),
\end{equation}
valid up to a global phase that plays no role in the dynamics. The blockade condition is enforced by the projectors $\hat{P}_{\langle j \rangle}$ and $\hat{Q}_{\langle j \rangle} = \mathds{1}_{\langle j \rangle} - \hat{P}_{\langle j \rangle}$, where $\hat{P}_{\langle j \rangle}$ projects onto the subspace in which none of the nearest neighbors of site $j$ are excited, and $\hat{Q}_{\langle j \rangle}$ ensures that excitations on neighboring qubits prevent transitions. Together, they implement the pseudo-Rydberg-blockade mechanism. The operator $\hat{\mathbb{U}}_{\chi_\ell,j}$ in Eq.~\eqref{eq:Uell} represents the local unitary evolution of the $j$-th qubit of type $\chi_\ell$ induced by $\hat{H}^{\rm eff}_{\rm drive,\ell}$:
\begin{equation}
\hat{\mathbb{U}}_{\chi_\ell,j} = \exp \Big\{-i \frac{\Omega_{\chi_\ell} \tau_{\chi_\ell}}{2} \big[ \cos(\phi_{\chi_\ell})\hat{\sigma}^{x}_j + \sin(\phi_{\chi_\ell})\hat{\sigma}^{y}_j \big] \Big\},
\label{Umathbb}
\end{equation}
where $(\Omega_{\chi_\ell}, \phi_{\chi_\ell})$ are the constant control parameters within $\tau_{\chi_\ell}$. Introducing the dimensionless parameters
\begin{align}
& \theta_\ell = \Omega_{\chi_\ell} \tau_{\chi_\ell}, \\
& \vec{n}_{\ell} = \big(\cos(\phi_{\chi_\ell}), \sin(\phi_{\chi_\ell}), 0 \big)^{\intercal},
\end{align}
the operator $\hat{\mathbb{U}}_{\chi_\ell, j}$ takes the compact form:
\begin{equation}
\hat{\mathbb{U}}_{\chi_\ell, j} = \exp \left[-i\frac{\theta_\ell}{2}\, \vec{\hat{\sigma}}_j \cdot \vec{n}_{\ell} \right],
\end{equation}
where $\vec{\hat{\sigma}}_j = \big( \hat{\sigma}^{x}_j, \hat{\sigma}^{y}_j, \hat{\sigma}^{z}_j \big)$ is the vector of Pauli matrices at site $j$.

Recalling that crossed qubits are driven at twice the Rabi frequency of regular ones, we decompose $\hat{U}_{\chi_\ell}$ as:
\begin{align}
\hat{U}_{\chi_\ell} &= \prod_{j \in \chi_{\ell}^{\rm r}} \Big(\mathds{1}_j \otimes \hat{Q}_{\langle j \rangle} + \hat{\mathbb{R}}_j(\theta_\ell,\vec{n}_{\ell}) \otimes \hat{P}_{\langle j \rangle} \Big) \nonumber \\
&\quad \times \prod_{j \in \chi_{\ell}^{\times}} \Big(\mathds{1}_j \otimes \hat{Q}_{\langle j \rangle} + \hat{\mathbb{R}}_j(2\theta_\ell,\vec{n}_\ell) \otimes \hat{P}_{\langle j \rangle} \Big),
\end{align}
where $\chi_\ell^{\rm r}$ and $\chi_\ell^{\times}$ denote the regular and crossed subsets of $\chi_\ell$-type qubits, and $\hat{\mathbb{R}}_j(\theta_\ell,\vec{n}_\ell)$ is the single-qubit rotation operator by angle $\theta_\ell$ around the Bloch sphere unit vector $\vec{n}_\ell$:
\begin{align}
\hat{\mathbb{R}}_j(\theta_\ell,\vec{n}_\ell) &= \exp \left[-\frac{i\theta_\ell}{2}\, \vec{n}_\ell \cdot \vec{\hat{\sigma}}_j \right] \\
&= \cos \!\left(\frac{\theta_\ell}{2}\right) \mathds{1}_j - i\sin \!\left(\frac{\theta_\ell}{2}\right) \vec{n}_\ell \cdot \vec{\hat{\sigma}}_j.
\end{align}
Finally, $\hat{U}_{\chi_\ell}$ can be written in the condensed notation:
\begin{equation}
\hat{U}_{\chi_\ell} = \hat{W}_{\chi_\ell}(\theta_\ell, \vec{n}_\ell;\, 2\theta_\ell, \vec{n}_\ell),
\end{equation}
defined through:
\begin{align}
\label{WTRANSF}
&\hat{W}_{\chi_\ell}(\theta_\ell,\vec{n}_\ell;\,\theta_\ell',\vec{n}_\ell') = \hat{W}_{\chi_\ell^{\rm r}}(\theta_\ell,\vec{n}_\ell)\, \hat{W}_{\chi_\ell^{\times}}(\theta_\ell',\vec{n}_\ell'), \\
&\hat{W}_{\chi_\ell^\xi}(\theta_\ell,\vec{n}_\ell) = \prod_{j \in \chi_{\ell}^{\xi}} \Big(\mathds{1}_j \otimes \hat{Q}_{\langle j \rangle} + \hat{\mathbb{R}}_j(\theta_\ell,\vec{n}_\ell) \otimes \hat{P}_{\langle j \rangle} \Big),
\end{align}
with $\xi \in \{\rm r, \times\}$. In summary, each pulsed unitary $\hat{U}_{\chi_\ell}$ in $\hat{U}_{\rm tot}$ implements a controlled unitary gate of the form:
\begin{equation}
\label{fdfsdf}
\hat{U}_{\chi_\ell} = \hat{W}_{\chi_\ell^{\rm r}}(\theta_\ell,\vec{n}_\ell)\, \hat{W}_{\chi_\ell^{\times}}(2\theta_\ell,\vec{n}_\ell),
\end{equation}
acting simultaneously on $\chi_\ell^{\rm r}$ and $\chi_\ell^{\times}$ qubits and inducing single-qubit rotations by angles $\theta_\ell$ and $2\theta_\ell$, respectively, around a unit vector $\vec{n}_\ell$ lying in the $xy$-plane (i.e.\ $\vec{z} \cdot \vec{n}_\ell = 0$).

\subsection{Quantum information flow}

The quantum information flow is described by the shift unitary
\begin{equation}
\label{equshiftrev}
\hat{U}_{\rm shift} = \hat{\Pi}_{A^{\rm r}}\hat{\Pi}_{B}\hat{\Pi}_{C}\hat{\Pi}_{A^{\rm r}},
\end{equation}
where $\hat{\Pi}_{\chi}$ denotes global pulses on all $\chi^\xi$-type qubits with $\xi \in \{\rm r, \times\}$ and $\chi \in \mathcal{S}$. The explicit pulse sequences are as follows.
The operator $\hat{\Pi}_{B}$ is decomposed as $\hat{\Pi}_{B} = \hat{U}^{(4)}_B\hat{U}^{(3)}_{B}\hat{U}^{(2)}_{B}\hat{U}^{(1)}_{B}$, with each step defined by:
(a) $\hat{U}^{(1)}_{B}$: phase $\phi_B^{(1)} = 0$, duration $\tau_1 = \frac{3\pi}{4}\Omega_B^{-1}$;
(b) $\hat{U}^{(2)}_{B}$: phase $\phi_B^{(2)} = \pi/2$, duration $\tau_2 = \pi\Omega_B^{-1}$;
(c) $\hat{U}^{(3)}_{B}$: phase $\phi_B^{(3)} = \pi$, duration $\tau_3 = \frac{\pi}{4}\Omega_B^{-1}$;
(d) $\hat{U}^{(4)}_{B}$: phase $\phi_B^{(4)} = -\pi/2$, duration $\tau_4 = \pi\Omega_B^{-1}$,
where $\Omega_B$ is the Rabi frequency of the regular $B$-type qubits. An analogous sequence applies to $\hat{\Pi}_{C}$.

For $\hat{\Pi}_{A^{\rm r}}$, which induces $\pi$-pulses on the regular $A$-type qubits only, the four-step decomposition $\hat{\Pi}_{A^{\rm r}} = \hat{U}^{(4)}_{A}\hat{U}^{(3)}_{A}\hat{U}^{(2)}_{A}\hat{U}^{(1)}_{A}$ reads:
(a) $\hat{U}^{(1)}_{A}$: phase $\phi_A^{(1)} = 0$, duration $\tau_1 = \frac{\pi}{2}\Omega_A^{-1}$;
(b) $\hat{U}^{(2)}_{A}$: phase $\phi_A^{(2)} = \pi/2$, duration $\tau_2 = \pi\Omega_A^{-1}$;
(c) $\hat{U}^{(3)}_{A}$: phase $\phi_A^{(3)} = \pi$, duration $\tau_3 = \frac{\pi}{2}\Omega_A^{-1}$;
(d) $\hat{U}^{(4)}_{A}$: phase $\phi_A^{(4)} = -\pi/2$, duration $\tau_4 = \pi\Omega_A^{-1}$,
where $\Omega_A$ is the Rabi frequency of the regular $A$-type qubits.

\subsection{One- and two-qubit gates}

Whenever the ICC is located in a $\chi$-type column with $\chi \in \{B,C\}$, single-qubit gates are realized by composing sequences of the form
\begin{eqnarray}
\hat{U}_{\chi,1} &=& \hat{W}_{A}(\pi, \vec{z},\, 0, \vec{u})\, \hat{W}_{\chi}(0, \vec{u},\, \theta/2, -\vec{n}) \nonumber \\
&&\times\, \hat{W}_{A}(\pi, \vec{z},\, 0, \vec{u})\, \hat{W}_{\chi}(0, \vec{u},\, \theta/2, \vec{n}),
\end{eqnarray}
where the latter implements a single-qubit gate on the $\chi$-type crossed element, $\vec{n}$ is a unit vector in the $xy$-plane ($\vec{n} \cdot \vec{z} = 0$), $\vec{u}$ is an arbitrary unit vector, and $\theta \in [0, 2\pi)$. Using Eq.~\eqref{WTRANSF}, this simplifies to:
\begin{equation}
\hat{U}_{\chi,1} = \hat{Z}_{A^{\rm r}}\, \hat{W}_{\chi^{\times}}(\theta/2,-\vec{n})\, \hat{Z}_{A^{\rm r}}\, \hat{W}_{\chi^{\times}}(\theta/2,\vec{n}),
\end{equation}
where $\hat{Z}_{A^{\rm r}} = \hat{W}_{A^{\rm r}}(\pi, \vec{z}) \propto \hat{\sigma}^z$ applies a $\hat{\sigma}^z$ gate to all the $A^{\rm r}$-type qubits. Since $\chi$-type crossed qubits are three columns apart, $\hat{U}_{\chi,1}$ effectively applies the single-qubit rotation
\begin{equation}
\hat{\mathbb{R}}(\theta,\vec{n}) = \hat{\sigma}^{z}\, \hat{\mathbb{R}}(\theta/2,-\vec{n})\, \hat{\sigma}^{z}\, \hat{\mathbb{R}}(\theta/2,\vec{n})
\end{equation}
to the crossed element of the ICC when acting on $\ket{\mathcal{W}^{(N)}_R(\psi)}$. This property also holds when the $\chi$-type polarized state $\ket{g}^{\otimes N}$ is replaced by a generic $N$-qubit state for the ICC.

For two-qubit gates, the ICC must first be aligned with the relevant gate column, which relies on the $A^{\times}$-type crossed qubits (couplers). A single evolution under:
\begin{equation}
\hat{U}_{A,2} = \hat{Z}_{A^{\times}} = \hat{W}_{A}(0, \vec{u},\, \pi, \vec{z}) = \hat{W}_{A^{\times}}(\pi, \vec{z}),
\end{equation}
which involves $(\theta_\ell, \theta_\ell') = (0,\pi)$ and leaves all $A^{\rm r}$-type qubits invariant, induces a conditional-phase shift~\cite{Nielsen2010} on the crossed qubits, thereby realizing two-qubit entangling operations.

\section{Derivation of the RF+RWA Lindblad master equation}
\label{rf_jwa_LME}

We show analytically that, for the relaxation and dephasing noise model, the dissipator is left invariant by the RF+RWA transformation. Consequently, starting from the LME with the original Hamiltonian $\hat{H}(t)$, defined in Eqs.~\eqref{H0text} and~\eqref{Hdrive}, and then applying the transformation is exactly equivalent to working with the LME governed by the RF+RWA Hamiltonian of Eqs.~\eqref{Hfinal} and~\eqref{hdrive2-main}, with the dissipator unchanged. In other words, the RF+RWA approximations and the dissipative terms commute.

Both frames appear explicitly in this Appendix. Consistently with the main text, $\hat{\rho}(t)$ always denotes the density matrix in the rotating frame---the object propagated in Eq.~\eqref{LMEeff} and in all the simulations---whereas $\hat{\rho}^{\rm lab}(t)$ denotes its laboratory-frame counterpart, the two being related by $\hat{\rho}(t) = \tilde{U}(t)\,\hat{\rho}^{\rm lab}(t)\,\tilde{U}^{\dag}(t)$. A tilde on an operator, $\tilde{O}(t) := \tilde{U}(t)\hat{O}(t)\tilde{U}^{\dag}(t)$, likewise denotes its rotating-frame counterpart.

We begin with the LME in the laboratory frame:
\begin{align}
\label{LMElab}
\frac{\mathrm{d}}{\mathrm{d}t}\hat{\rho}^{\rm lab} &= -\frac{i}{\hbar}\big[\hat{H}(t),\hat{\rho}^{\rm lab}\big] \\ &+ \sum_{j,\alpha} \Big( \hat{L}_{j,\alpha} \hat{\rho}^{\rm lab} \hat{L}^{\dag}_{j, \alpha} - \frac{1}{2} \big\{ \hat{L}^{\dag}_{j, \alpha} \hat{L}_{j,\alpha}, \hat{\rho}^{\rm lab}\big\}\Big). \nonumber
\end{align}

\paragraph{Rotating frame.}
We first apply the rotating frame (RF) transformation. Writing $\hat{\rho}^{\rm lab}(t) = \tilde{U}^{\dag}(t)\hat{\rho}(t)\tilde{U}(t)$, substituting into Eq.~\eqref{LMElab} and using the unitarity of $\tilde{U}(t)$, one finds:
\begin{align}
\frac{\mathrm{d}}{\mathrm{d}t}\hat{\rho} = &-\frac{i}{\hbar}\Big[\tilde{H}(t),\hat{\rho}\Big] + \sum_{j,\alpha} \Big(\tilde{L}_{j,\alpha} \hat{\rho} \tilde{L}^{\dag}_{j, \alpha} - \frac{1}{2} \big\{ \tilde{L}^{\dag}_{j, \alpha} \tilde{L}_{j,\alpha}, \hat{\rho}\big\}\Big) \nonumber \\ 
&-\hat{\rho}\left[\partial_t\tilde{U}(t)\right]\tilde{U}^{\dag}(t) -\tilde{U}(t)\left[\partial_t\tilde{U}^{\dag}(t)\right]\hat{\rho} .
\end{align}
Since $\tilde{U}(t)$ is unitary, $\tilde{U}^{\dag}(t)\tilde{U}(t) = \mathds{1}$ implies $\tilde{U}(t)\partial_t\tilde{U}^{\dag}(t) = -[\partial_t\tilde{U}(t)]\tilde{U}^{\dag}(t)$, so that the last two terms combine into a single commutator and the LME simplifies to:
\begin{align}
\frac{\mathrm{d}}{\mathrm{d}t}\hat{\rho} = &-\frac{i}{\hbar}\Big[\tilde{H}(t) + i\hbar\big(\partial_t \tilde{U}(t)\big)\tilde{U}^{\dag}(t),\hat{\rho}\Big] \nonumber \\ 
&+ \sum_{j,\alpha} \Big(\tilde{L}_{j,\alpha} \hat{\rho} \tilde{L}^{\dag}_{j, \alpha} - \frac{1}{2} \big\{ \tilde{L}^{\dag}_{j, \alpha} \tilde{L}_{j,\alpha}, \hat{\rho}\big\}\Big).
\end{align}
For the transformation $\tilde{U}(t)$ of Sec.~\ref{sec:model} the counter-term evaluates to $i\hbar\big[\partial_t\tilde{U}(t)\big]\tilde{U}^{\dag}(t) = -\sum_{\chi \in \mathcal{S}}\sum_{j \in \chi}(\hbar\omega_{\mathrm{d},\chi}/2)\,\hat{\sigma}^z_j$, which shifts the local qubit energies from $\omega_\chi$ to the detunings $\omega_\chi' = \omega_\chi - \omega_{\mathrm{d},\chi}$ and thus produces $\hat{H}_0^{\mathrm{eff}}$ of Eq.~\eqref{Hfinal}. Discarding the fast counter-rotating terms then yields the full effective Hamiltonian $\hat{H}^{\mathrm{eff}}(t) = \hat{H}_0^{\mathrm{eff}} + \hat{H}_{\mathrm{drive}}^{\mathrm{eff}}(t)$ of Eqs.~\eqref{Hfinal} and~\eqref{hdrive2-main}.

The transformed jump operators are $\tilde{L}_{j,\alpha} = \tilde{U}(t)\hat{L}_{j,\alpha}\tilde{U}^{\dag}(t)$. The key observation is that both $\hat{\sigma}^z_j$ and $\hat{\sigma}^-_j$ are \emph{eigenoperators} of the rotating-frame generator. Using
\begin{align}
\label{eigenop}
&e^{i\theta\hat{\sigma}^z/2}\,\hat{\sigma}^-\,e^{-i\theta\hat{\sigma}^z/2} = e^{-i\theta}\,\hat{\sigma}^-, \\
&e^{i\theta\hat{\sigma}^z/2}\,\hat{\sigma}^z\,e^{-i\theta\hat{\sigma}^z/2} = \hat{\sigma}^z,
\end{align}
which follow from $[\hat{\sigma}^z,\hat{\sigma}^-] = -2\hat{\sigma}^-$ and $[\hat{\sigma}^z,\hat{\sigma}^z]=0$, the rotated jump operators acquire at most a global phase:
\begin{align}
\label{Ltilde_minus}
&\tilde{L}_{j,-} = e^{-i\omega_{\mathrm{d},\chi}t}\,\hat{L}_{j,-} = \sqrt{\gamma_-}\,e^{-i\omega_{\mathrm{d},\chi}t}\,\hat{\sigma}^-_j, \\&\tilde{L}_{j,z} = \hat{L}_{j,z} = \sqrt{\gamma_z}\,\hat{\sigma}^z_j.
\end{align}
This phase is immaterial in the dissipator, since each jump operator appears together with its conjugate. For the relaxation channel,
\begin{align}
\tilde{L}_{j,-}\,\hat{\rho}\,\tilde{L}^{\dag}_{j,-} &= \gamma_-\,\hat{\sigma}^-_j\,\hat{\rho}\,\hat{\sigma}^+_j,
\end{align}
and likewise $\tilde{L}^{\dag}_{j,-}\tilde{L}_{j,-} = \gamma_-\,\hat{\sigma}^+_j\hat{\sigma}^-_j = \hat{L}^{\dag}_{j,-}\hat{L}_{j,-}$. The dephasing channel is trivially invariant since $\tilde{L}_{j,z} = \hat{L}_{j,z}$. We emphasize that, unlike the bit-flip operator $\hat{\sigma}^x = \hat{\sigma}^+ + \hat{\sigma}^-$, whose two components rotate with opposite phases and thus mix $\hat{\sigma}^x$ and $\hat{\sigma}^y$ in the rotating frame, the lowering operator $\hat{\sigma}^-$ transforms with a single phase and is therefore left invariant \emph{exactly}---no time averaging or RWA is required at the level of the dissipator.

Collecting all terms, and noting that the RWA applies only to the coherent (Hamiltonian) part, the dissipative dynamics of the ladder architecture within the RF+RWA framework is governed precisely by Eq.~\eqref{LMEeff} of the main text, with $\alpha\in\{-,z\}$ and the jump operators $\hat{L}_{j,-} = \sqrt{\gamma_-}\,\hat{\sigma}^-_j$, $\hat{L}_{j,z} = \sqrt{\gamma_z}\,\hat{\sigma}^z_j$ unchanged with respect to the laboratory frame. We note that, in contrast with the dephasing channel for which $(\hat{\sigma}^z_j)^2 = \mathds{1}_j$ produces a constant loss term, the relaxation loss term involves the projector $\hat{\sigma}^+_j\hat{\sigma}^-_j = \ket{e_j}\bra{e_j}$ and therefore does not reduce to a $\hat{\rho}$-proportional constant; the dissipator of Eq.~\eqref{LMEeff} is thus kept in full Lindblad form.

\section{Numerical methods}
\label{sec:methods}

Each numerical result presented in the main text is obtained using the quantum trajectory method, also known as the quantum jump method~\cite{dalibard1992,dum1992,molmer1993,gardiner1992,carmichael1993,daley2014}, within a tensor network framework. This approach reformulates the Lindblad master equation \eqref{LMEeff} as an ensemble average over stochastic pure-state evolutions under the Markovian approximation. For each trajectory, the time-evolved quantum state propagates under a non-Hermitian effective Hamiltonian
$\hat{H}^{\mathrm{sim}}(t) = \hat{H}^{\mathrm{eff}}(t) - (i\hbar/2)\sum_{j,\alpha}\hat{L}_{j,\alpha}^{\dag}\hat{L}_{j,\alpha}$,
punctuated by random quantum jumps in which the Lindblad operators $\hat{L}_{j,\alpha}$ are applied stochastically to the time-evolved state. At each time step $\delta t$, a jump of type $\alpha$ occurs at site $j$ with probability $\delta p_{j,\alpha} = \delta t\,\langle\Psi(t)|\hat{L}^{\dag}_{j,\alpha}\hat{L}_{j,\alpha}|\Psi(t)\rangle$, in which case the state is updated as $\hat{L}_{j,\alpha}\ket{\Psi(t)}$ and renormalized; otherwise the non-Hermitian evolution proceeds and the state is renormalized to compensate the norm deficit~\cite{daley2014}. The expectation value of any observable $\hat{O}$ at time $t$ is then recovered by averaging over $N_{\mathrm{traj}}$ independent trajectories:
$O(t) = N_{\mathrm{traj}}^{-1}\sum_{k=1}^{N_{\mathrm{traj}}} \langle \Psi^{(k)}(t) | \hat{O} | \Psi^{(k)}(t) \rangle$,
where $\ket{\Psi^{(k)}(t)}$ denotes the $k$-th trajectory realization. The main computational advantage of this approach is that each trajectory involves the propagation of a pure state rather than a full density matrix, reducing the Hilbert-space scaling from $\mathcal{O}(d^2)$ to $\mathcal{O}(d)$.

This pure-state propagation combines naturally with the matrix product state (MPS) representation~\cite{schollwock2005,schollwock2011}, which provides an efficient compression of weakly entangled many-body quantum states. The 2D ladder system, coupled through nearest-neighbor antiferromagnetic interactions, is mapped onto a 1D quantum lattice model with long-range interactions that are non-algebraic and break translational invariance; this mapping is efficient precisely because the number of inter-row couplers is small (only $N-1$ for $N$ rows). Within the MPS/MPO framework, the entanglement structure of the state is analyzed at each step and contributions below a threshold singular value are truncated, yielding a compact and controlled representation of the many-body state. This approach is particularly effective for the low-entanglement states characteristic of low-dimensional quantum systems.

Time propagation within each trajectory is performed using the time-dependent variational principle (TDVP)~\cite{haegeman2011} augmented with ancillary Krylov subspace (AKS) expansion~\cite{yang2020}. This combined approach finds the optimal path within the MPS manifold at fixed bond dimension while dynamically enlarging the bond dimension via global Krylov vectors, preserving the unitarity of coherent evolution steps. Simulations were carried out using the Julia implementation of the ITensor package~\cite{itensor,itensor2} on computer clusters, with parallelization over independent quantum trajectories. In all cases, the convergence of the results was verified through systematic analysis of the numerical cutoffs: number of trajectories $N_{\mathrm{traj}} = 2000$, maximum MPS bond dimension $\max(m) = 200$, and MPS truncation threshold $\lambda = 10^{-12}$.
\end{document}